\documentclass[letterpaper]{article} 
\usepackage{aaai2026}  
\usepackage{times}  
\usepackage{helvet}  
\usepackage{courier}  
\usepackage[hyphens]{url}  
\usepackage{graphicx} 
\usepackage{natbib}  
\usepackage{caption} 
\usepackage{amssymb}

\usepackage{booktabs}          
\usepackage[table]{xcolor}     

\usepackage{csquotes}              
 
\usepackage{tabularx}
\usepackage{placeins}
\usepackage{float}
 
\usepackage{soul}
\definecolor{stathl}{RGB}{244,212,200}
\sethlcolor{stathl}

\usepackage[most]{tcolorbox}
 
\newtcolorbox{participantquote}[1][]{%
  colback=gray!4, colframe=gray!50, boxrule=0.4pt,
  left=8pt, right=8pt, top=6pt, bottom=6pt,
  fontupper=\itshape, breakable, #1
}

\usepackage[colorinlistoftodos, textsize=small]{todonotes}
\usepackage{verbatim}              

\title{Auditing AI-Washing in German Startups:
A Mixed-Methods Study of Marketing Claims and Perceptions}

\author{
    A Pranav, Luke Hartmann, Anne Lauscher
}
\affiliations{
    Trustworthy AI Lab\\
    University of Hamburg, Germany
}

\begin{document}

\maketitle

\begin{abstract}
AI-washing is the use of `AI' as a marketing term where products' claims are exaggerated, likely to capture the hype from investors and consumers.
In this paper, we examine how startups market AI and how people perceive those claims.
We audited the German startup market, applying an annotation codebook of seven dimensions to 100 startups and conducting 63 expert interviews.
Startups exaggerated AI capability in three forms: claims without inspectable evidence, performance numbers without methodology, and claims without acknowledged limits.
The capabilities they claimed rarely matched the underlying system, and consumers who had encountered exaggerated claims before withdrew trust from later AI marketing.
AI-washing has also been linked to the marketing of unethical products: 38\% of the sample were flagged for moderate or serious ethical concern, in domains such as surveillance and automated decisions in regulated fields.
Together these findings show how AI-washing happens: founders exaggerate capability to attract investors who reward AI branding they cannot verify, and consumers in unfamiliar domains take the marketing at face value.
We recommend that startup AI claims be brought under disclosure requirements, with the codebook offered to regulators, investors, and consumer-protection bodies for routine audits.
\end{abstract}
 
\section{Introduction}
\label{sec:Introduction}

Media coverage and investor attention have inflated what the term `artificial intelligence' promises, and startups increasingly build their products and pitches around it \citep{Floridi2024AIHypeBubble, Garud2025Hype}.
In Germany, 45.1\% of the 2025 startup cohort name AI as a core product component \citep{Hirschfeld2025StartupMonitor}.
Investors drive towards AI branding: founders who present their product as AI-driven raise capital more easily than those who do not, and funding often goes towards AI-based business models \citep{Rady2025StartupHype}.
Many consumers treat AI as a \emph{\enquote{near-magical capability}} rather than a bounded engineering system, and consumers with lower AI literacy overestimate what an advertised product can do \citep{cave2019hopes, tully2025lower}.
These conditions reward what we call \textbf{AI-washing}: \textit{the deceptive use of AI-related claims, in which a company exaggerates, misrepresents, or obscures the actual role of AI in its products and operations} \citep{Elhajjar2025AIwashing, kozarkiewicz2025artificial}.
 
In practice, deception in marketing happens in three ways: claiming an ability the system does not have, reselling another company's system as proprietary technology, and exaggerating a capability without saying what the system actually does \citep{aditya2001deceptive}.
Often, a customer or an investor cannot check what the advertised AI does behind a product, or whether it exists at all, because the marketing does not disclose the method and no independent test is available \citep{CFAInstitute2025AIWashing, Barrios2025AIDisclosures}.
Investors often pressure startups to brand their products as \textit{AI-powered} or \textit{smart}.
Behind that label, some of these products gather and model large amounts of personal data \citep{zuboff2015big, SadowskiPasquale2015}.
Such products often result in tracking, behavioural profiling, and automated decisions the user never knowingly agreed to, even as the marketing calls this personalisation or assistance \citep{susser2019technology, nissenbaum2004privacy}.
The AI label reframes these functions as product features, so users and regulators do not subject them to the scrutiny that surveillance and profiling would otherwise attract \citep{Mathur2019DarkPatterns, crawford2021atlas}.

We investigate the following question in this paper: \textbf{\textit{What does AI-washing look like in the marketing of German AI startups, what drives founders and marketers to produce it, and how do consumers and domain experts receive it?}}
We explore this question via a \textbf{scoping study} scores the marketing copy of 100 German AI startups on a seven-dimension codebook and a composite index, and 63 expert \textbf{interviews} (53 written, ten oral) examine why founders and marketers make these claims and how audiences receive them.
We report three findings:

\begin{enumerate}
    \item \textbf{Startups market capabilities current AI cannot deliver.} Startups advertised AI functions that exceed what current systems can reliably do, and applied AI to tasks a simpler approach would handle; most described their AI only in buzzwords without saying what it did, and many gave AI more marketing weight than its role in the product that it actually has.
    \item \textbf{Consumers are getting more skeptical on exaggerated AI-washed claims.} Startups overstated capabilities by advertising claims with no publicly inspectable evidence (43\%), displaying performance numbers with no disclosed methodology, and capability claims that acknowledge no limits (76\%). Consumers who had encountered exaggerated claims before reported lower trust in \enquote{AI-powered} claims.
    \item \textbf{AI-washing enables ethical harm by marketing high-risk applications as ordinary AI products.} Some of the startups openly marketed surveillance, autonomous decisions in regulated domains, and behavioural tracking and manipulation as ordinary AI products. 38\% of the sample were flagged for moderate or serious ethical concern (19\% serious), all in categories the EU AI Act classifies as high-risk or prohibits outright.
\end{enumerate}

\paragraph{Related work.}
Even before AI, startups have long exaggerated what their products can do to convince investors and customers that the venture is ahead of its market \citep{Zott2007SymbolicMgmt, Rutherford2009LegitimacyLie}.
Venture capital rewards this behaviour, because valuations follow what investors anticipate a technology to become rather than what the product can demonstrably do \citep{Kampmann2024VCFetish}.
Studies of AI-washing document this exaggeration in forms that range from inflated capability claims \citep{Nyilasy2025AIwashingTA, Barrios2025AIDisclosures, Elsayed2026AIWashingLegitimacy} to third-party technology resold as proprietary AI \citep{Elhajjar2025AIwashing, kozarkiewicz2025artificial}.
\citet{Zhang2026CapabilitiesDiscourses} analysed the capability discourses of large AI companies and argued that such claims persuade by reinstating hegemonic ideas about society and identity.

\section{Methods}

We designed a mixed-methods study to explore three sides of AI-washing in the German startup market: the supply side of public marketing claims, the motivations behind those claims, and the demand side of audience perception.
We combine two methods, each described in turn below: a scoping study of marketing copy, and expert interviews on motivation and audience perception.
For the scoping study, we annotated 100 German AI startups against a seven-dimension codebook ($D_1$--$D_7$) that scores how a startup describes its AI in marketing copy and what the underlying product actually does, and aggregated the per-dimension scores into a composite AI-washing index.
To explain why these claims are made and how audiences perceive them, we conducted 63 expert interviews with researchers, marketers, investors, and employees of AI startups: 53 in writing and ten orally in semi-structured sessions.

\subsection{Scoping study}\label{sec:scoping}

\subsubsection{Sample}\label{sec:sample}
We constructed the sample from two publicly accessible directories of German startups: the Startup-Atlas\footnote{\url{https://www.startup-atlas.de/startups} (Accessed in early 2026)} and the AI Startup Landscape maintained by the appliedAI Institute\footnote{\url{https://www.appliedai-institute.de/ressourcen/ai-startups-deutschland/} (Accessed in early 2026)}.
We filtered the Startup-Atlas by the IT and Communication domain and the Artificial Intelligence tag, and drew a random subsample from the AI Startup Landscape to reduce selection bias.
Inclusion required a German registered office, an AI claim in public marketing, a functional website with substantive content at the time of annotation, and pre-IPO (Initial Public Offering) startup status.
Applying these criteria yielded a final sample of 100 startups.

The first two authors served as annotators.
Both have combined backgrounds in business administration, marketing, AI, and AI ethics, matching the codebook's mix of technical and communication dimensions.

\subsubsection{Codebook}\label{sec:codebook}
We developed the codebook through iterative open coding rather than top-down specification, drawing on \citet{Birhane2022ForgottenMargins} for the dual-annotation protocol and on \citet{Bommasani2024FMTI} and \citet{Liesenfeld2024OpenWashing} for multi-dimensional company scoring.
The development workflow ran in five steps:
\begin{itemize}
    \item Both annotators independently reviewed approximately 30 startups without a predefined scheme and recorded qualitative observations on AI claims, marketing language, technical transparency, evidence of functionality, and ethical implications.
    \item Comparing the two sets of observations surfaced five recurring patterns: (i) \enquote{AI} used as a branding label with no detectable AI functionality, (ii) third-party APIs (most often the OpenAI API) repackaged as proprietary products, (iii) quantitative performance claims made without methodology or source, (iv) headline marketing that promised more than the technical descriptions on product subpages supported, and (v) ethically loaded applications obscured behind positive AI framing.
    \item We formalised these patterns into seven coding dimensions ($D_1$--$D_7$), each with anchored rating scales and boundary-case decision rules.
    \item Both annotators dual-coded a 10-startup calibration set and revised anchor wording for cases that produced disagreement.
    \item Each annotator then independently coded 50 startups, with a separate 10-startup overlap set dual-coded; Cohen's $\kappa$ (macro-averaged) on this set was $0.82$. 
\end{itemize}

We treated a product as containing AI where its advertised function would use learned models, including large language models, and not where it runs on fixed rules, keyword matching, or scripted workflows.
The substance dimension ($D_2$) then grades where that AI comes from: proprietary models at the top of the scale, adapted or aggregated third-party models in the middle, a single wrapped API near the bottom, and no detectable AI at the floor.

After the seven dimensions were coded, each startup received a composite AI-washing score from 1 to 5 ($\kappa = 0.86$ on the score).
The composite is a single overall judgment anchored at 1 (transparent, proportionate, evidenced, and ethically sound) and 5 (misrepresentation across several dimensions, in some cases with ethical obfuscation).
Ethical concern ($D_6$) raises the composite where AI-washing language normalises a problematic application, and strong functionality evidence ($D_3$) or disclosed limitations ($D_7$) lower it for an otherwise concerning profile; annotators verified the provisional composite against these adjustments before finalising.

Table~\ref{tab:codebook_summary} summarises the seven dimensions.
We additionally recorded, by manual inspection, whether each website carried text or imagery showing signs of AI generation.
We did not use automated AI-content detectors, because current detectors are not reliable enough for evidentiary use \citep{Sadasivan2024Detect}.
This observation supports later qualitative discussion but does not feed the composite index.

\begin{table}[t]
  \centering
  \small
    \begin{tabularx}{\linewidth}{l l X}
    \toprule
    \textbf{Dim.} & \textbf{Scale} & \textbf{One-line definition} \\
    \midrule
    $D_1$ & 1--5 & \textbf{AI Claim Specificity}: how concretely the startup describes what its AI does, from buzzword to fully disclosed model and method. \\
    $D_2$ & 1--5 & \textbf{AI Capability Authenticity}: technical substance behind the AI label, from no detectable AI to proprietary models with disclosed architecture. \\
    $D_3$ & 1--5 & \textbf{Evidence of AI Functionality}: what an external observer can verify, from no evidence to independently validated benchmarks. \\
    $D_4$ & 0--3 & \textbf{Quantitative Claim Substantiation}: whether numeric performance claims are backed by source or methodology; 0 indicates no quantitative claim was made. \\
    $D_5$ & 1--5 & \textbf{Marketing Proportionality}: match between the prominence of AI in marketing and its role in the product. \\
    $D_6$ & 0--3 & \textbf{Ethical Concern Level}: ethical risk of the AI application itself, coded independently of marketing behaviour. \\
    $D_7$ & 1--4 & \textbf{Transparency About Limitations}: whether the startup discloses what its AI cannot do, including accuracy bounds and human-in-the-loop dependencies. \\
    \bottomrule
  \end{tabularx}
\caption{Codebook summary.}
  \label{tab:codebook_summary}
\end{table}

For the Ethical Concern dimension ($D_6$), annotators scored from 0 (no apparent concern) through 1 (minor: potential for misuse or minor privacy implications), 2 (moderate: personal-data processing or behavioural tracking), to 3 (serious: surveillance, emotion detection, discriminatory potential, or an application that meets a high-risk category of the EU AI Act).
A score of 3 was grounded in the EU AI Act's \citep{EUAI2024Act} high-risk definitions, and the ethical subcategory (surveillance, privacy, manipulation, discrimination, lack of human oversight, or vulnerable persons) was recorded alongside the score.

\subsubsection{Annotation procedure}\label{sec:procedure}
For each startup, the annotator followed a multi-source workflow that took roughly one hour to complete:
\begin{itemize}
    \item Read the homepage and priority subpages (\texttt{/about}, \texttt{/technology}, \texttt{/how-it-works}, or local equivalents).
    \item Interacted with the product where a free trial or live demo was available, to verify whether the marketed AI behaviour was observable rather than merely described.
    \item Reviewed company and founder activity on LinkedIn, X, and other public social media for product descriptions, founder background, and shifts in framing over time.
    \item Consulted external sources (Crunchbase, Tracxn, investor press releases) for funding stage, team size, and prior product positioning.
    \item Scored each dimension only against the codebook anchors and only on the basis of publicly visible material; coded N/A with a reason where a dimension could not be scored.
\end{itemize}

Each annotation also recorded fixed metadata for later cross-sectional analysis: sector, product type, business model, founding year, location, funding stage, team size, founder background, and whether \enquote{AI} appears in the company name.

Beyond scores and metadata, several free-text fields recorded substantive qualitative evidence about each startup:
\begin{itemize}
    \item \textbf{AI-washing hints}: we annotated exaggeration signals lifted verbatim from each website, including superlative capability claims, vague autonomy or intelligence boasts, and unsubstantiated quantitative comparisons. 
    \item \textbf{Ethical concerns}: we annotated the type of application when it met the EU AI Act's high-risk criteria, including surveillance, biometric categorisation, credit assessment, and behavioural manipulation, regardless of how the marketing dimensions were scored.
    \item \textbf{Value proposition and product description}: we annotated the headline value proposition as advertised on the homepage alongside a sober description of what the product actually does.
    \item Finally, we documented edge cases and scoring difficulties that fed into periodic codebook refinement.
\end{itemize}

\subsubsection{Sample composition}\label{sec:composition}
The 100 startups divide by business model into 77 Business-to-Business (B2B), 15 mixed Business-to-Business and Business-to-Consumer (B2B/B2C), 7 Business-to-Consumer (B2C), and 1 Business-to-Business-to-Consumer (B2B2C).
Sectors span finance, e-commerce, healthcare, marketing technology, legal technology, entertainment, industrial automation, logistics, and human resources.
Product types break down into 49 software-as-a-service tools, 30 platforms, 14 hardware-software combinations, 5 consultancies, and 2 API products.
Founding years range from 2014 to 2025; 50 of the 100 startups were founded in or after 2022.
Berlin (29) and Munich (19) form the largest geographic nodes, followed by Hamburg (8) and Stuttgart (5).

\subsubsection{Definitions}
In this paper, we use the following conventions.
$\rho(D_X, D_Y)$ denotes the Spearman rank correlation between dimensions $D_X$ and $D_Y$, and $\rho(D_X, \text{composite})$ the correlation between a dimension and the composite score, computed over the full sample of 100 startups unless a smaller $n$ is reported.
Dimension scales run in different directions (a low $D_5$ but a high $D_6$ both indicate greater severity), so for correlations with the composite we align each dimension so that higher values indicate greater AI-washing severity; a positive $\rho$ therefore always means that greater severity on the dimension accompanies a higher composite score.
Mann-Whitney $U$ denotes a two-group rank comparison, which we use to compare annotation scores between founder-background groups and Likert ratings between interview-participant groups.
All $p$-values are two-sided, and all group comparisons are exploratory.
We write $s \in \{4, 5\}$ to state that a score $s$ takes one of the values listed in the braces.
The \textit{high-composite} group denotes the 39 startups with composite scores $\in \{4, 5\}$; all other distributional statistics use the full sample unless stated otherwise.

\subsection{Interviews}
\label{sec:interviews}
 
The 53 written interviews followed a structured question set that recorded how people encounter and react to AI marketing claims.
We sought participants working in software development, sales and marketing, or management to document the internal pressures behind those claims.
The instrument collected demographic information, self-rated AI familiarity (1--5), four 5-point Likert items, two binary items on encounters with AI-washing, and four open-text prompts: an open definition of AI, examples of startup AI claims, an account of any personal AI-washing experience, and a closing comment on AI marketing.
The two binary items asked whether the participant had encountered German startups advertising AI capabilities and whether they had personally encountered what they would consider AI-washing as a consumer or at work; the latter item defines the encountered-versus-not grouping used in the analysis of the written interviews.
The first Likert item measured reaction to the label \enquote{AI-powered} (1=very sceptical, 5=very trustful).
The remaining three asked whether current AI is capable of reliably and ethically deploying surveillance, of making complex autonomous decisions, and of taking over the role of therapists or psychologists.
 
\paragraph{Recruitment and language.}
We recruited participants through social media such as LinkedIn and Bluesky.
Recruitment calls went out in both English and German, and we ran two language versions of the instrument in parallel.
German free-text responses were translated into English before analysis.
We obtained 53 complete responses.
 
\paragraph{Sample composition.}
The 53 written-interview participants are AI-literate professionals and graduate students.
68\% self-rated their AI familiarity at 4 or 5 out of 5, 77\% held a Bachelor's or Master's degree, and 55\% fell in the 25--34 age bracket.
47\% work or study in technical computing (computer science, NLP, machine learning, data analytics), 17\% in business, finance, or marketing, and 36\% in journalism, media, regulation, health care, and other domain-specialist roles.
This is not a probability sample of German consumers, and we do not treat it as one.
The findings describe how AI-literate professionals, the audience best equipped to check an AI claim against the underlying technology, react to AI marketing.
 
\paragraph{Oral interviews.}
We recruited ten additional participants through purposive sampling whose roles span sales, marketing, programming, and management in AI startups, drawing on the authors' professional networks.
Interviews ran 30--60 minutes online by video call.
Each session was audio-recorded with the participant's informed consent and transcribed verbatim.
Three interviews conducted in German were translated into English (manually, by a native speaker) and spot-checked against the original transcripts.
The interview guide opened with background and AI-familiarity questions and then moved into perceptions of AI startups, first-hand encounters with AI-washing, organisational drivers of AI marketing claims (investor pressure, internal visibility incentives, regulatory gaps), and views on remediation.
 
\paragraph{Analysis.}
For the written interviews, we report per-item distributions, means, and the proportion of responses in the lower two and upper two Likert categories.
After transcribing each oral interview, the first and second authors independently open-coded the transcripts \citep{strauss1998basics}.
Over three weeks, these authors met to discuss shared patterns across the interview codes and written responses.
Throughout, we cite interview participants as P1--P63.

\paragraph{Limitations.}
The annotation sample covers only German startups, so the findings do not generalise to other jurisdictions.
Annotation are based on the marketing copies, public demos, and free trials where available, rather than on long-term hands-on product use, so scores largely reflect what firms chose to show.
The interview sample is small, so it cannot show what a full-population sample would: lower AI literacy predicts greater receptivity to AI claims \citep{tully2025lower}, so how general consumers receive the same claims remains open.

\section{Finding 1: Startups Market Capabilities Current AI Cannot Deliver}
\label{subsec:finding1-overclaim}
 
Startups in the high-composite group marketed AI capabilities that current AI systems cannot reliably deliver.
Often startups applied AI to product tasks a simpler, deterministic approach would handle more reliably.
59\% of the startups described their AI only in buzzwords or superficial terms, without detail about what the AI did [$D_1 \in \{1,2\}$].
35\% had no actual AI, or relied on an LLM (Large Language Model) wrapper without proprietary contribution [$D_2 \in \{1,2\}$].
40\% gave AI more marketing weight than its role in the product warranted [$D_5 \in \{1,2\}$].
Startups whose AI was based on an LLM wrapper were significantly more likely to overstate AI's role in their product ($\rho(D_2, D_5) = +0.51$, $p < 0.001$, $n = 99$, the strongest pairwise correlation in our codebook).
In table~\ref{tab:f1_receipts} we show verbatim public-facing claims from six startups in this group.
 
\begin{table}[t]
\centering
\small
\begin{tabularx}{\linewidth}{@{}>{\raggedright\arraybackslash}X@{}}
\toprule
\textbf{A legal-research startup.} \emph{\enquote{We develop AI without hallucinations: our AI invents nothing.}}~\emph{[transl.]} \\
\midrule
\textbf{A digital-transformation advisory startup.} \emph{\enquote{Our AI provides solutions \& best practices across your entire company, enabling smarter decisions without workshops, consultants or IT projects.}} \\
\midrule
\textbf{A bio-robotic surveillance startup.} \emph{\enquote{We release cockroaches equipped with a custom-built backpack for control, sensing, and secure communication, enabling precise navigation and real-time data collection.}} \\
\midrule
\textbf{A call-centre voice-AI startup.} \emph{\enquote{Digital call center workers are trusted by enterprises to deliver the same or better customer satisfaction as human agents. At scale.}} \\
\midrule
\textbf{A specialty-chemicals formulation startup.} \emph{\enquote{We develop AI Twins for complex chemical formulations and materials based on our Deep Knowledge Integration approach.}} \\
\midrule
\textbf{An automotive voice-AI startup.} \emph{\enquote{We develop Digital Employees that could think, talk, and act alongside their human counterparts.}} \\
\bottomrule
\end{tabularx}
\caption{Examples of marketing copy showing the exaggerated claims (composite $\geq 4$).
German quotes are translated and marked \emph{[transl.]}.}
\label{tab:f1_receipts}
\end{table}

\paragraph{Buzzword AI labels without a description of the AI.}
Startups marketed their AI under labels such as \enquote{AI credit assessment}, \enquote{real-time AI for the contact centre}, \enquote{agentic banking}, and \enquote{intelligent autonomy}.
None said which part of the product was the AI, or how it differed from rule-based automation or an off-the-shelf model.
31\% of the startups carried AI descriptions the annotator coded as unclear or technically unspecific.
Startups with business-only founders were significantly more likely to overstate AI capabilities than those with technical founders (Mann-Whitney $U = 287$, $p = 0.003$).
They also marketed AI more prominently than its actual role in the product warranted ($U = 76$, $p = 0.002$).
 
P12, a written-interview participant, described their experience of encountering this type of AI-washing:
\begin{quote}
\emph{\enquote{Potential business partner claimed their tool would reduce workload by providing \textbf{AI powered solution} to assess documents submitted in risk management context. Was basically only keyword search, still required close human supervision.}}~(P12)
\end{quote}
P42 saw the same pattern at consumer-goods scale, where AI labels attach to products whose functions predate the label:
\begin{quote}
\emph{\enquote{I often notice that products are marketed as AI-powered, even in areas where it makes absolutely no sense (e.g., Axe shower gel, Bosch washing machine). I think this suggests that the products are \enquote{intelligent}, even though the functions existed before the AI hype.}}~\emph{[transl.]}~(P42)
\end{quote}
Once the buyer reads the AI label as persuasion, the claim loses its standing as a product fact \citep{friestad1994persuasion}.
 
\paragraph{AI applied to tasks that do not call for AI.}
We observed startups marketing AI for tasks AI is the wrong tool for, and in some cases for tasks a deterministic system would handle more reliably.
Such use cases are \textit{technological solutionism}: the assumption that the right answer to a problem must involve technology, even where simpler engineering or human practice would do \citep{morozov2013save, selbst2019fairness}.
One startup advertises an outcome (no hallucinations, full source traceability) achievable by deterministic retrieval over the same source corpus but not by current LLMs; another reframes consultancy work as a self-serve AI platform whose AI label substitutes for the work (Table~\ref{tab:f1_receipts}).
P31, an AI researcher among the written-interview participants, described from the user side:
\begin{quote}
\emph{\enquote{It is very frustrating when generative AI is shoehorned into applications that it is not ready for, and thus the failure rate in using the product increases exponentially. I am not against AI in principle (I research it), but it is nascent in many of the contexts where it is currently deployed.}}~(P31)
\end{quote}
Current LLMs cannot guarantee zero hallucinations, and an AI label on a consultancy would likely provide homogenised advice.
 
\paragraph{Wrappers and human labour sold as proprietary AI ($D_2$).}
60\% of startups marketed AI prominently without owning a proprietary AI system [$D_2 \in \{1,2,3\}$].
72\% of the high-composite group lacked proprietary AI: they had no AI at all, a third-party wrapper, or a combination of existing AI tools (28 of 39).

A LinkedIn-growth tool advertises \emph{\enquote{Leverage the power of AI for LinkedIn Growth}} in the main text, whereas a footnote on the same page says the product is \emph{\enquote{powered by OpenAI's ChatGPT 4 technology}} ($D_2 = 1$, $D_5 = 1$).
Startups have been building wrappers around closed-source LLMs rather than engineering systems with fine-tuned models of their own.
P2, an AI researcher tracking the German startup scene, reaches the same conclusion across many such websites:
\begin{quote}
\emph{\enquote{In the end I tend to come to the conclusion: okay, that's just a nice interface with an API key to ChatGPT, a nice function. It gets copied and done. It's just a quick cash grab for a year or so.}}~\emph{[transl.]}~(P2)
\end{quote}
P62 recognised the same wrapper pattern from the consumer side:
\begin{quote}
\emph{\enquote{Often, for specific cases, a prompt is simply forwarded to ChatGPT. You think it is a custom, trained model, but in the end it is just a ChatGPT context that offers no unique selling point.}}~\emph{[transl.]}~(P62)
\end{quote}

P4 described a case where a no-code app-builder had marketed an AI assistant capable of producing finished client apps from a single prompt at a fraction of normal development cost.
When the company collapsed publicly in 2025, reporting confirmed that outsourced developers had done much of the underlying coding \citep{Orosz2025}.
When a company sells human work as AI, the workers behind the system become invisible to customers, and their labour falls outside the wage and protection norms that would apply if the same work were sold as software development \citep{gray2019ghost, casilli2025waiting}.
In both cases, what is sold as AI is something else: a third-party model the startup does not own, or human labour standing in for the advertised AI.
 
\paragraph{The AI label is attached to products whose function does not match the claim.}
At a former employer of P6, a sustainability-reporting workflow was relabelled as \enquote{AI reporting} although the underlying work was data aggregation and dashboarding (P6, paraphrased).
P5 reported an advertised capability that did not work in use:
\begin{quote}
\emph{\enquote{I cannot dial any phone number that contains a 2, because the system does not understand the way I say 2.
I have never managed it.
But it is advertised prominently: voice control, AI-driven, and so on.
Why do we need this kind of AI when it is biased against German accents?}}~\emph{[transl.]}~(P5)
\end{quote}
The automotive voice-AI startup in Table~\ref{tab:f1_receipts} promises \emph{\enquote{Digital Employees}} that can think, which no current voice system can do [$D_1 = 2$, $D_2 = 3$, $D_5 = 2$].
In each case the advertised AI is either absent behind a relabelled non-AI workflow or unable to perform the advertised function.

\section{Finding 2: Startups Exaggerate AI Capabilities And Customers Are Losing Trust}
\label{subsec:finding2-overclaiming}

Finding~1 documented startups claiming AI capabilities their products do not have.
Startups produce the discrepancy by exaggerating: presenting an AI capability with greater scope, certainty, or sophistication than the underlying technology supports.
\textbf{Three forms of exaggeration recur across the scoping sample: capability claims with no inspectable evidence behind them, numerical claims that name no methodology, and capability claims that acknowledge no limits.}
Disproportionate marketing weight ($D_5$), the mismatch between AI prominence in marketing copy and the actual role of AI in the product, was the strongest correlate of the composite AI-washing score ($\rho(D_5, \text{composite}) = +0.57$, $p < 0.001$).
The interviews trace these patterns to financing and visibility incentives, and participants who had previously encountered exaggerated claims reported lower trust in subsequent AI marketing.
Table~\ref{tab:finding2-receipts} shows verbatim marketing copy from five exemplar startups.

\begin{table}[t]
\centering
\small
\begin{tabularx}{\linewidth}{>{\raggedright\arraybackslash}X}
\toprule
\textbf{An AI-agent startup.} \emph{\enquote{We develop The Autonomous Agent that gives you superpowers.}} \\
\midrule
\textbf{A manufacturing-optimisation startup.} \emph{\enquote{Our AI saves companies up to 96\,\% of production costs.}} \\
\midrule
\textbf{An autonomous-retail startup.} \emph{\enquote{$10\times$ lower theft than the industry average, with the highest prosecution rates.}} \\
\midrule
\textbf{A field-sales-AI startup.} \emph{\enquote{Thinks ahead, learns from corrections, and never drops the ball.}} \\
\midrule
\textbf{A knowledge-management startup.} \emph{\enquote{Your employees can ask anything and get a perfect, validated answer instantly.}} \\
\bottomrule
\end{tabularx}
\caption{Verbatim marketing quotes from five exemplar startups demonstrating exaggerated claims.}
\label{tab:finding2-receipts}
\end{table}

\paragraph{Capability claims with no inspectable evidence.}
43\% of the startups claimed AI capabilities without publicly inspectable evidence behind them [$D_3 \in \{1,2\}$], and only 2\% offered evidence at the highest tier of the codebook.
We found that most websites offering any evidence presented it as client logos, anonymised case studies, or controlled product demos.
None of these forms lets an outside reader audit the claimed AI capability, because the seller controls what the artefact shows: a logo proves a customer relationship rather than what the AI did for that customer, an anonymised case study cannot be checked with the source firm, and a controlled demo runs on inputs the seller selected.
The startup in Row~1 of Table~\ref{tab:finding2-receipts} names its product as an \emph{\enquote{Autonomous Agent}} that confers \emph{\enquote{superpowers}} on the user, without specifying what task the agent performs autonomously, what the superpowers consist of operationally, or how either claim could be verified.
P16 reported the same absence of checkable content in the claims they had encountered: \emph{\enquote{It is often left to the reader's imagination. Specific examples refer to data volumes or speed.}}~(P16).
By offering self-controlled artefacts in place of independent verification, these startups foreclose the basic check a stakeholder needs to evaluate the product.

\paragraph{Numerical claims with no methodology.}
Among startups that make any quantitative claim, methodology is almost never disclosed: 79\% of quantitative claimants (49 of 62) show no substantiation for their claims ($D_4 = 1$), and none are fully transparent ($D_4 = 3$).
The high-composite group follows the same pattern: 95\% make either no quantitative claims or only unsubstantiated ones.
Among startups that make quantitative claims, weaker substantiation accompanied higher AI-washing severity ($\rho(D_4, \text{composite}) = +0.25$, $p = 0.05$, $n=62$).
The most common form is a multiplier or \emph{\enquote{up to X\,\%}} phrasing with no baseline, no customer reference, and no methodology.
Rows~2 and~3 of Table~\ref{tab:finding2-receipts} illustrate two examples within the pattern.
The manufacturing-optimisation firm advertises a 96\% saving with no comparison baseline disclosed; the autonomous-retail firm names a baseline (\emph{\enquote{the industry average}}) but does not define which industry, which measurement, or which time window.
P33 traces the practice to capital pressure: \emph{\enquote{AI companies are desperately seeking use cases so they can justify to their funders the capital that they have invested. Because of this, they try to make their AI approaches seems~[\textit{sic}] far more sophisticated and `human' than they are.}}~(P33).
Without disclosed baselines or methodology, the numbers function as branding: they invite comparison but provide nothing to compare against.

\paragraph{Exaggeration by hiding limitations.}
76\% of the startups named no edge cases, accuracy bounds, data dependencies, or human-in-the-loop requirements anywhere in their marketing material [$D_7 \in \{1,2\}$], and only 7\% reached the top disclosure tier [$D_7 = 4$].
Hiding limits works the same way as the low claim specificity ($D_1$) and disproportionate marketing weight ($D_5$) covered earlier: each makes a system sound more capable than it is.
Two startups go further and promise that the AI never fails: a field-sales AI \emph{\enquote{thinks ahead, learns from corrections, and never drops the ball}}; a knowledge platform invites employees to ask \emph{\enquote{anything}} and receive a \emph{\enquote{perfect, validated answer instantly}} (Table~\ref{tab:finding2-receipts}, Rows~4 and~5).
Neither line names a failure mode, an accuracy bound, a domain boundary, or a human review step.
Hiding limits is a baseline practice even outside the high-composite group, where 72\% of startups disclosed no limitations, and the high-composite group disclosed still less (82\%; $\rho(D_7, \text{composite}) = +0.24$, $p < 0.05$).
P21 described this from their former employment, where a product marketed as fully automated depended on a continuous human review team:
\begin{quote}
\emph{\enquote{The classification system of cybersecurity alerts I was working with was presented to clients as being fully capable of delivering automatic alerts.
In reality, a whole team of humans was working each day to analyze, correct, refine and package results.
I discovered the discrepancy as I was working in the company, both with the \enquote{refining team} and with the commercial team selling the product as fully automated to clients.
To my knowledge, no specific consequences happened to the company.}}~(P21)
\end{quote}
What P21 saw is the inside view of the pattern in Rows~4 and~5: capabilities advertised as perfect did have limitations the marketing did not disclose.
P9 noted that academic conferences have begun to require an explicit limitations paragraph as a structural mitigation against the same pattern in published research.

\paragraph{AI-generated marketing material as a correlated signal.}
We observed apparent AI-generated images or text on 13\% of the sites.
We flagged sites whose copy carried the lexical signatures of LLM-assisted writing \citep{kobak2024delving}, or whose avatars and product images looked synthetic.
We observed that the use of AI-generated marketing material is correlated with AI-washing severity: 77\% of the flagged sites (10 of 13) sit in the high-composite group, against 33\% of the 87 unflagged sites ($\rho(\text{AI-gen flag}, \text{composite}) = +0.25$, $p = 0.011$).
LLM-generated text tends toward superlatives and infallibility framing, the same pattern observed in the exaggerated copy in Table~\ref{tab:finding2-receipts}, which may explain the correlation.

\paragraph{The mechanism of exaggeration.}
Participants trace the exaggeration to two pressures: an external one from investors and public funders, and an internal one from management and marketing on the people building the product.
P2, who works in German startup financing, connects the AI label to fundraising: \emph{\enquote{AI is currently the buzzword for financing. That's why everyone does it.}}~(P2).
Participants who worked inside the firms report the internal pressure.
At P3's firm, the instruction came from management: \emph{\enquote{We were really caught up in the hype, where the word was: AI has to go in everywhere.}}~(P3).
P4, on the marketing side, saw the website claim run ahead of the product: while a small development team was still in its first three planning meetings on an AI feature, the company website already described the firm as \emph{\enquote{AI First}}, presenting a prototype as a deployed product.
P6 gives the budget logic behind such labels: top management \emph{\enquote{would rather put extra funding to any project which has the name AI because they believe that it might bring in huge innovation}}~(P6), so employees who want funding for a project label it AI.
P7 recalls the same renaming with an earlier buzzword: companies added \emph{\enquote{Blockchain}} to their names to attract investment without changing their products, and AMD now sells its processor line as \emph{\enquote{Ryzen AI}}.
Across the interviews, the AI claim comes from people who do not build the product: founders speaking to investors, top management, and marketing.
The developers closest to the technical work (P1, P4) do not decide how the product is presented.

\paragraph{Demand-side recognition and trust withdrawal.}
In the written interviews, we examined how participants perceive exaggeration.
Participants reported personally encountering AI-washing as consumers or in their work (53\%).
On the trust-in-\emph{\enquote{AI-powered}} item, participants who had encountered AI-washing were markedly more sceptical than those who had not (encountered: median~1, mean~1.71; not encountered: median~3, mean~2.48; Mann-Whitney $U = 190$, $p = 0.003$).
The two groups did not differ in self-rated AI familiarity ($U = 448$, $p = 0.069$).
Participants distrust \enquote{AI-powered} claims after they have seen exaggeration, not because they know more about AI.
P18 saw the pattern at two employers before graduate school:
\begin{quote}
\emph{\enquote{Over the course of my career, AI capabilities have always been systematically overclaimed -- the benefits often oversold, and the costs understated.
At both companies I was at pre-grad school, AI was often presented in ways that led with the potential, rather than the actual truth of the product being provided.
AI tools were decontextualized of the training data and attendant limitations, and performance was reported in ways that did not take into account generalization to end-user usecases.}}~(P18)
\end{quote}
Consumers can detect exaggeration, and once they have, they trust later \enquote{AI-powered} claims less.
\citet{friestad1994persuasion} call this \textit{persuasion knowledge}: a consumer who recognises a persuasion attempt assigns the message a different meaning.
\emph{Finding~3} examines startups whose overselling reaches domains where the application itself raises ethical concern: surveillance, autonomous decision-making, and behavioural manipulation.

\section{Finding 3: AI-washing Leads To Ethical Risks}
\label{subsec:finding3-ethical}
 
Startups in the high-composite group were more likely to market applications that raised ethical concerns ($\rho(D_6, \text{composite}) = +0.41$, $p < 0.001$).
19\% of the 100 startups marketed applications raising serious ethical concerns [$D_6 = 3$], and a further 19\% marketed applications raising moderate concerns [$D_6 = 2$].
All of these applications fall within categories the EU AI Act classifies as high-risk or prohibits outright~\citep{EUAI2024Act}.
The three subcategories with the largest counts were surveillance, autonomous decisions in regulated domains, and behavioural tracking and manipulation.
Table~\ref{tab:finding3-receipts} reproduces marketing copy from five exemplar startups in the $D_6 = 3$ cluster.
 
\begin{table}[t]
\centering
\small
\begin{tabularx}{\linewidth}{>{\raggedright\arraybackslash}X}
\toprule
\textbf{A geospatial-AI startup} \emph{(Surveillance)} \\
\emph{\enquote{Our AI can detect unusual activities, monitor crowd movements, and identify security threats, aiding law enforcement and security agencies in maintaining public order and safety.}} \\
\midrule
\textbf{A parking-surveillance startup} \emph{(Surveillance)} \\
\emph{\enquote{Persons, animals, and personal data are tracked using our in-house intelligent software.}} \\
\midrule
\textbf{A defence bio-robotics startup} \emph{(Surveillance)} \\
\emph{\enquote{Deploy silent sensors and conduct covert surveillance in dense, unstable, and politically sensitive urban environments.}} \\
\midrule
\textbf{A voice-AI call-centre startup} \emph{(Manipulation)} \\
\emph{\enquote{We use LLMs, human-like voices and background call center music to deliver \textgreater{}90\,\% CSAT score.}} \\
\midrule
\textbf{An AI fertility-clinic startup} \emph{(Autonomous decisions)} \\
\emph{\enquote{We use our computer vision-based tool to select your highest-quality embryo based on comprehensive clinical and lab data.}} \\
\bottomrule
\end{tabularx}
\caption{Examples of marketing copy from startups with serious ethical concerns ($D_6 = 3$).}
\label{tab:finding3-receipts}
\end{table}

\paragraph{Surveillance.}
Three of the five exemplar startups market surveillance applications.
The EU AI Act classifies law-enforcement and public-space monitoring as high-risk under Annex~III §6 of Regulation~2024/1689 and additionally restricts real-time remote biometric identification under Article~5(1)(h).
Municipalities market these systems as \enquote{smart city} projects; the term often covers surveillance rather than urban infrastructure and keeps public scepticism away \citep{SadowskiPasquale2015, Kitchin2014RealTimeCity}.
Facial recognition, the technology behind many of these systems, misidentifies people at rates that differ by factors of ten or more across racial and sex categories~\citep{Buolamwini2018GenderShades, Grother2019NISTFRVT}.
Many such systems train on person re-identification datasets whose subjects never consented to collection~\citep{Peng2021DatasetStewardship}.
As in Finding~1, the startups advertise capabilities current AI cannot deliver.

Written-interview participants doubted that current AI can be deployed reliably or ethically for surveillance.
70\% rated capability in the lower two scores of a five-point scale (mean $2.17$).
Participants trained in computer science, AI, NLP, or business informatics rated capability lower still, at $1.70$ on average against $2.45$ for the rest of the sample (Mann-Whitney $U = 197$, $p = 0.011$).

P1 worked inside an AI-surveillance startup that sold camera-based footfall analysis to shopping malls; the firm pitched capabilities its cameras could not support:
\begin{quote}
\emph{\enquote{We were mainly AI systems behind it.
I never understood the point of doing retail analytics with surveillance.
We cannot even determine their age or gender via surveillance cameras.
These cameras are really bad and they try to get your video from really far away, so you cannot really do that kind of stuff.
But we were trying to oversell to the investors that we can do that kind of stuff.
We were trying to oversell them like `okay, we can figure out who is doing robbery in your mall,' but that's not really possible to do that.}}~(P1)
\end{quote}

Asked whether AI surveillance can currently be deployed ethically and reliably, P10 said it can be neither:
\begin{quote}
\emph{\enquote{Of course not ethically and reliably.
Firstly, it does not work well.
AI based surveillance have a lot of false positives as they work on low-quality security cameras.
So even on a responsible level do you want to deploy AI when it does not work well at all?
And also what kind of knowledge will they gain from it?
It's just a huge data mess.}}~(P10)
\end{quote}

In P1's account, only people working inside the firm could see that the marketing did not match the engineering.
Buyers (police forces, security agencies, parking operators) cannot test such claims before deployment, and the people the cameras record are never asked at all.
An exaggerated surveillance product therefore harms people who never bought it.

\paragraph{Autonomous decisions in regulated domains.}
Two of the five exemplar startups market AI for decisions normally made by a regulated professional: an AI fertility-clinic startup with an embryo-selection tool (the fifth exemplar in Table~\ref{tab:finding3-receipts}), and a startup that offers autism screening for children and an AI therapy platform.
The fertility-clinic startup claims its AI makes \enquote{highest-quality} embryo selections.
However, the leading commercial system in this class did not outperform trained embryologists in a randomised clinical trial~\citep{Illingworth2024EmbryoRCT}.

Written-interview participants doubted that current AI can make decisions of this kind.
57\% of participants rated AI as not capable of such decisions and 32\% rated it capable (mean $2.57$).
Participants trained in computer science, AI, NLP, or business informatics gave a mean rating of $2.15$ against $2.82$ for the rest of the sample (Mann-Whitney $U = 230$, $p = 0.058$); as on the surveillance item, the technically trained group was the more sceptical one.

P23 rated AI as fully capable on this item and still opposed the use:
\begin{quote}
\emph{\enquote{I think AI is capable of making decisions like medical and legal ones but I don't think it should.}}~(P23)
\end{quote}
P1, asked in the oral interview about complex decision-making, also opposed it:
\begin{quote}
\emph{\enquote{I definitely don't think that AI should be used to do complex decision making processes.
I don't think that AI should be used for, let's say, crime and policing stuff, giving out loans, or determining if you have cancer or not.
\ldots Even if you try to properly train your AI, AI will have biases.
AI does not provide proper reasoning.}}~(P1)
\end{quote}

The second startup screens children for autism spectrum disorder from voice recordings and pairs neurodivergent families with therapists through an LLM platform.
The screening falls under Annex~III §2 as a medical-device safety component; the therapy platform under Annex~III §5 as access to an essential service.
Participants were most sceptical on the therapy item: 81\% placed AI in the lower two scores in the written interviews (mean $1.85$).
Current language models lack the contextual and situational grounding clinical practice requires, and audits report risks of privacy violation, sycophancy, persuasion, overtrust, and emotional dependence~\citep{Moore2025LLMTherapists}.

P17, who reported conscious daily use of generative AI, worried most about crisis scenarios:
\begin{quote}
\emph{\enquote{AI cannot replace the specifically human element of interaction.
This is why claims for products like AI therapists greatly worry me.
Yes, there is anecdotal evidence that some people feel better talking to them in the short term, but AI is in no way equipped to handle and direct a person in a crisis scenario.}}~(P17)
\end{quote}

\paragraph{Privacy issues and tracking.}
Startups present manipulation, behavioural tracking, and AI-generated outputs built on personal data collected without consent as product features (Table~\ref{tab:finding3-receipts}).
The call-centre product places automated outbound calls in which the synthesised voice and dialogue are AI-generated, with no disclosure to the recipient who answers.
\citet{Matz2024GenerativePersuasion} found that LLM messages personalised to a recipient were more persuasive than generic ones across consumer, political, and health domains.
A behavioural-analytics product reads visitor behaviour during a web session and reshapes the page in real time to lift conversion, with no disclosure to the visitor; \citet{Mathur2019DarkPatterns} catalogued interface designs of this kind as dark patterns across 11{,}000 shopping sites.
\citet{Susser2019OnlineManipulation} call this online manipulation: hidden influence on a person's choices, made cheap to deploy at scale by digital infrastructure.

P5, an investor and AI-strategy adviser, encountered undisclosed data collection in person at a German trade fair, where a startup offered passers-by a \enquote{psychological assessment} generated from biometric input collected on the spot.
P5 asked the vendor how the system was supposed to work:
\begin{quote}
\emph{\enquote{There was one person saying: do you want a psychological assessment of yourself? I said, how is that supposed to work? You have to put your finger here. Your eye has to look in here, and then I have to put my hand somewhere else. Then I said: so you want all my biometric data and you give me some kind of three-page printout.}}~(P5)
\end{quote}
The visitor hands over fingerprint, iris, and palm data on the strength of an AI label, and receives a printout for which no scientific basis exists \citep{stark2021physiognomic, barrett2019emotional}.
The pitch describes the assessment and says nothing about what happens to the collected data.

P34 worked barista shifts at a multinational coffee chain that deployed two AI systems on the shop floor, a computer-vision inventory counter and a workforce-planning model:
\begin{quote}
\emph{\enquote{Shift supervisors had to increasingly use AI technologies for tasks that shouldn't have necessarily required AI.
\ldots they had to use AI to plan the layout of the store during our peak rush, a model that often miscalculated where people should go and led to strain on certain baristas who had to do more work than others.
\ldots they had to use a computer vision system to do the stock inventory for them as it was supposed to make counting inventory quicker.
In reality, it just made it more frustrating and longer due to the amount of false positives.}}~(P34)
\end{quote}
Both systems underperformed against the marketed claim, and the underperformance made the shifts harder: the planning model dictated where each barista worked during peak rush and loaded some with more work than others, and the vision system's false positives put the counting back on the staff.
Sold as efficiency tools, the systems in practice placed workers under algorithmic decisions and monitoring during every shift, and collected tracking data about them.

AI-washing in all three cases conceals the regulatory class to which the underlying application belongs.
The EU AI Act places each application in a risk category: high-risk biometric identification, regulated medical device, prohibited manipulation.
None of this appears in the marketing that customers and investors read.
Findings~1 and~2 showed AI-washing costing buyers money and trust; here, startups sell systems the Act treats as high-risk or prohibited, and the harm falls on the people these systems watch.

\section{Discussion}
\label{sec:discussion}

This paper documented AI-washing in German startups across three positions: the marketing copy on startup websites, the accounts of experts who advise and invest in the market, and the responses of consumers who buy AI products.
German AI startups marketed products as autonomous or proprietary AI systems when those products wrapped third-party models, ran rule-based logic, or contained no AI at all (\emph{Finding~1}).
Consumers who had encountered exaggerated claims trusted subsequent AI marketing less; startups lose the trust of the consumers who detect the exaggeration (\emph{Finding~2}).
Startups also marketed surveillance, autonomous decisions in regulated domains, and behavioural tracking and manipulation as ordinary AI products, so these applications reached customers and investors with no mention of the high-risk or prohibited status the EU AI Act assigns them (\emph{Finding~3}).
We provide recommendations for policymakers and auditors that target the marketing copy rather than the systems behind it, because AI-washing is marketing that claims more than the startup has built.

\paragraph{Recommendations.}
First, any company that commercially advertises AI capabilities should publish, on the product page, a short structured statement of what data the system collects about the people who use it and what the system cannot reliably do; datasheets \citep{gebru2021datasheets} and model cards \citep{mitchell2019model} already specify the content categories such a statement can reuse, and this content is the part marketing copy omits.
Second, because a disclosure statement is still self-reported, an independent body should certify commercial AI marketing claims by inspecting a system's technical documentation and performance benchmarks under confidentiality; the EU AI Act's conformity assessment already delegates a comparable inspection of high-risk systems to designated third parties \citep{EUAI2024Act}, and this proposal extends that assessment from the system to the marketing claim made about it.
Third, both measures require deciding which startups warrant scrutiny, and the codebook can serve as the first-pass screen: a regulator or investor can score a startup's public marketing copy without access to the underlying system, and escalate to system-level review only the startups whose marketing the screen flags.

\paragraph{Conclusion.}
This paper documents AI-washing in the German startup market through an annotation codebook applied to 100 startups and 63 expert interviews (53 written, ten oral).
We demonstrate the incentives that sustain AI-washing: founders exaggerate AI capability to attract investors who reward AI branding, investors fund products whose AI they cannot easily verify, and consumers in unfamiliar domains take the marketing at face value.
The pressure that drives exaggeration also pushes some startups into applications that raise ethical concern, in domains such as surveillance, automated medical and legal decisions, and persuasion aimed at vulnerable users, where an overstated capability claim risks direct harm rather than only misdirected funding.
AI-washing persists because the parties positioned to check the claims cannot currently do so at low cost: investors lack the technical detail, consumers lack the domain knowledge, and regulators lack the empirical basis.
We recommend that AI claims in startup marketing be brought under disclosure requirements and routine audit, and we offer the codebook as a low-cost instrument for regulators, investors, and consumer-protection bodies to apply that audit at population scale rather than case by case.

\section{Acknowledgments}
We thank our anonymous reviewers for their feedback.
We thank the participants of our study for sharing their perspectives with us.
The work of Trustworthy AI lab is funded under the Excellence Strategy of the German Federal Government and States.

\section{Ethics Statements}
\label{sec:ethics}

The interviews constitute human-subjects research, and their protocols were reviewed and approved by the ethics review board of the University of Hamburg Business School before recruitment began.

\paragraph{Positionality.}
The lead authors have backgrounds in machine learning, AI ethics, business administration, and marketing, and study the German startup market as residents of Germany working at a German university.
Before this study, we worked in startups and in marketing roles, and have written and approved the kind of promotional copy this paper audits.
We knew from that work that founders and marketers write capability claims under funding and management pressure, so the codebook scores what the public record shows and the interview guide asks about those pressures.
The same work taught us to expect exaggeration, and we may have scored ambiguous claims more harshly than annotators without industry experience would.
The anchored scales and boundary-case rules limit how far that expectation can move a score, and the interviews describe the same patterns independently of our annotation.

\paragraph{Annotation of public marketing materials.}
The scoping study coded publicly accessible marketing materials of registered German firms, and we did not collect or store any non-public data about these companies, their founders, or their employees.
We anonymise the 100 startups in the text, receipts tables, and figures, and identify each only by city, sector, and product category.
The anonymisation matches the claim we make: our findings describe how often AI-washing patterns recur across the sample, not whether any single named firm has misled a specific investor or customer; that is a legal question the dual-annotation protocol cannot settle.
We retain verbatim marketing quotes in the receipts tables because the wording is the analytic object and because a reader can verify the prevalence pattern by sampling the public directories.
We did not contact the startups for right-of-reply, because the material is already public and our claims concern population-level prevalence rather than the conduct of any single company, following prior population-level audits of public artefacts \citep{Birhane2022ForgottenMargins, Liesenfeld2024OpenWashing}.

\paragraph{Oral interviews.}
The ten oral-interview participants gave informed consent before each session; consent covered audio recording, verbatim transcription, anonymised use of quotations in academic publication, and the right to withdraw at any point up to manuscript submission.
We stored recordings on encrypted devices, deleted the audio files after transcription and a single accuracy check, and retained only the transcripts under participant codes with no identifying metadata.
Participants discussed current and prior employers, and we redacted firm identifiers and other distinguishing details from quoted excerpts where retention would re-identify a specific organisation or person.
We translated the three German-language sessions into English and spot-checked the translations against the original transcripts before quoting any passage.

\paragraph{Written interviews.}
Participants took part voluntarily and without compensation, and we recruited through open posts on LinkedIn and Bluesky with no targeted outreach to individuals.
The instrument collected no directly identifying information.
We quoted free-text responses under participant codes, after redacting verbatim references to current employers from the excerpts.
We ran the instrument in English and German in parallel and translated German free-text responses into English for analysis.

\bibliography{aaai2026}

\clearpage
\appendix

\section{Appendix A: Codebook Design}
\label{app:codebook}

This appendix expands the codebook summary in Table~\ref{tab:codebook_summary} into the full per-score anchors used by annotators.
Each dimension is presented in a single table with two columns: the score level and the anchor description that annotators applied to public-facing material.
Dimensions $D_1$--$D_5$ and $D_7$ are oriented so that higher scores denote stronger performance (more specific claims, more authentic AI, more evidence, more substantiated quantitative claims, more proportionate marketing, more transparent limitations).
$D_6$ is oriented so that higher scores denote greater ethical concern.
When in doubt between two adjacent anchors, annotators chose the lower-concern rating, so the composite is interpretable as a lower bound on AI-washing severity.

\begin{figure*}[h]
    \centering
    \includegraphics[width=\textwidth, height=0.85\textheight, keepaspectratio]{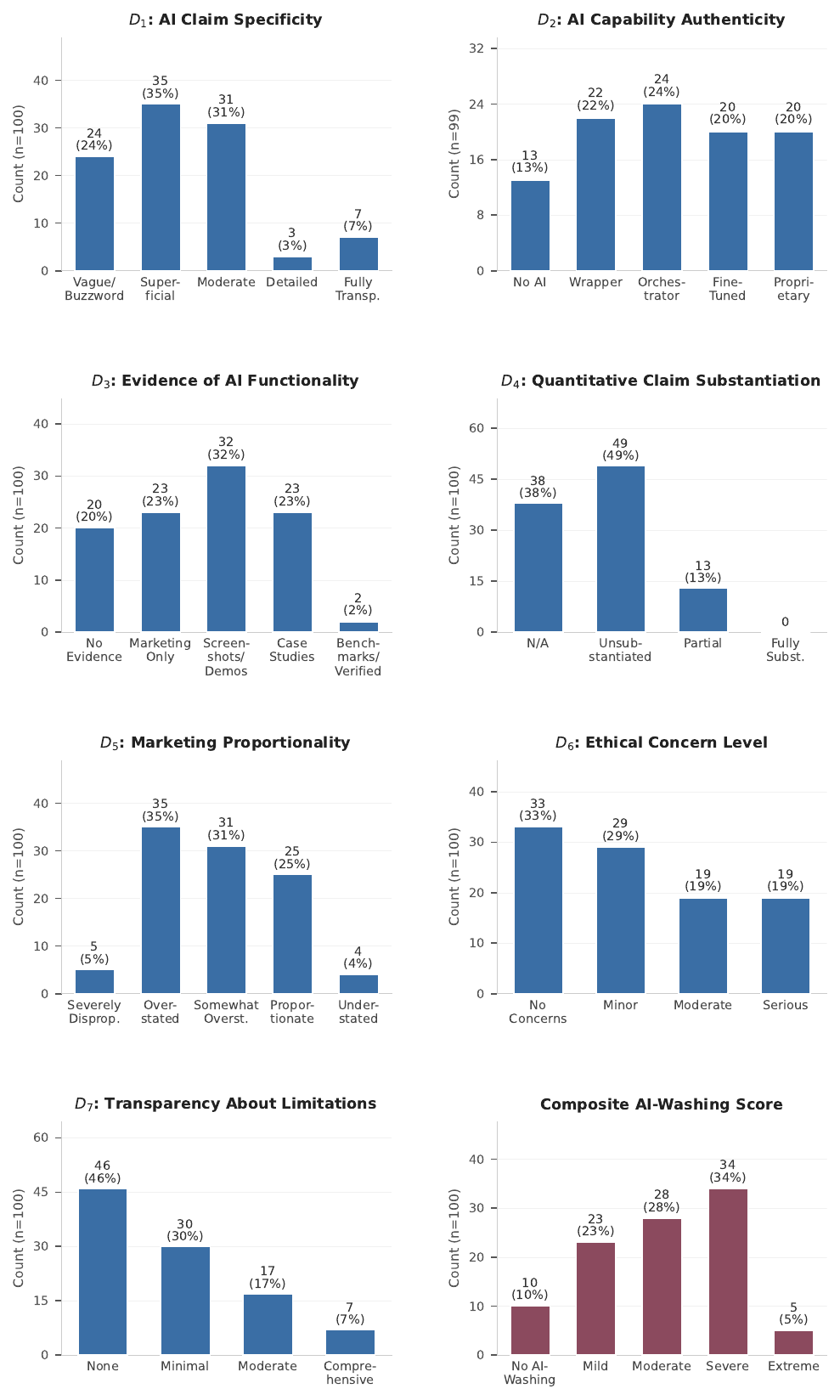}
    \caption{Score distributions on the seven codebook dimensions ($D_1$--$D_7$) and the composite AI-washing index.
    Each bar is labelled with the count and percentage of the sample.
    Scales differ by dimension: $D_1$, $D_2$, $D_3$, $D_5$ use 1--5; $D_4$ and $D_6$ use 0--3; $D_7$ uses 1--4; the composite uses 1--5.}
    \label{fig:app_dim_dist}
\end{figure*}

\paragraph{Dimension $D_1$: AI Claim Specificity (1--5).}
$D_1$ records how concretely a startup explains what its AI does, from a branding label with no functional content to a fully disclosed model and method.

\begin{table}[h]
\centering
\small
\caption{$D_1$: AI Claim Specificity anchors.}
\label{tab:d1_anchors}
\begin{tabularx}{\linewidth}{@{}c X@{}}
\toprule
\textbf{Score} & \textbf{Anchor description} \\
\midrule
1 & \textbf{Vague / Buzzword.} AI appears in headlines or branding only. No explanation of function. Examples: \enquote{Universal AI software} with nothing behind it; \enquote{Powered by AI} with zero detail. Generic subfield names (\enquote{NLP}, \enquote{machine learning}, \enquote{predictive analytics}) without any indication of how they apply to the product remain at this level. \\
\addlinespace
2 & \textbf{Superficial.} Some indication of what AI does (e.g., \enquote{AI-based analysis}) but no methods, models, or technical depth. \\
\addlinespace
3 & \textbf{Moderate.} General function explained (e.g., \enquote{uses NLP for text simplification}, \enquote{predictive maintenance}) but the description lacks specifics on models or methods. \\
\addlinespace
4 & \textbf{Detailed.} Clear explanation of techniques, model types, and integration (e.g., \enquote{object detection on aerial imagery using CNNs}). \\
\addlinespace
5 & \textbf{Fully Transparent.} Models named, methods described, data sources disclosed, with optional links to papers or repositories. \\
\bottomrule
\end{tabularx}
\end{table}

\paragraph{Dimension $D_2$: AI Capability Authenticity (1--5).}
$D_2$ records whether the AI behind the product is genuine and proprietary, repackaged from third-party APIs, or absent.

\begin{table}[h]
\centering
\small
\caption{$D_2$: AI Capability Authenticity anchors.}
\label{tab:d2_anchors}
\begin{tabularx}{\linewidth}{@{}c X@{}}
\toprule
\textbf{Score} & \textbf{Anchor description} \\
\midrule
1 & \textbf{No Detectable AI.} No visible AI component. AI is purely a marketing label, and the product could function without it. \\
\addlinespace
2 & \textbf{Third-Party Wrapper.} Uses OpenAI or another commercial API with minimal customisation. Signals include mentions of GPT or ChatGPT, \enquote{powered by OpenAI}, generic chatbot interfaces, and security text such as \enquote{we do not send your data to train third-party models}. \\
\addlinespace
3 & \textbf{Orchestrator.} Aggregates multiple existing AI tools into a unified platform without building new models. \\
\addlinespace
4 & \textbf{Fine-Tuned or Adapted.} Meaningfully customises existing models for a specific domain, with domain-specific training data, custom pipelines, or explicit fine-tuning claims. \\
\addlinespace
5 & \textbf{Proprietary AI.} Own models or algorithms with verifiable evidence: research papers, named architectures, benchmarks, public repositories, or patents. \\
\bottomrule
\end{tabularx}
\end{table}

\paragraph{Dimension $D_3$: Evidence of AI Functionality (1--5).}
$D_3$ records what an outside observer can verify about the marketed AI from publicly available material.

\begin{table}[h]
\centering
\small
\caption{$D_3$: Evidence of AI Functionality anchors.}
\label{tab:d3_anchors}
\begin{tabularx}{\linewidth}{@{}c X@{}}
\toprule
\textbf{Score} & \textbf{Anchor description} \\
\midrule
1 & \textbf{No Evidence.} No demos, screenshots, case studies, or benchmarks. \\
\addlinespace
2 & \textbf{Marketing Only.} Client logos, testimonials, generic success stories, or self-reported metrics without methodology. \\
\addlinespace
3 & \textbf{Screenshots / Demos.} Product screenshots, demo videos, interactive demos, or a free trial that shows AI behaviour in operation. \\
\addlinespace
4 & \textbf{Case Studies.} Published case studies with specific, attributed results (named clients, concrete outcomes). \\
\addlinespace
5 & \textbf{Benchmarks / Verified.} Technical benchmarks, peer-reviewed publications, open-source code, or independent validation. \\
\bottomrule
\end{tabularx}
\end{table}

\paragraph{Dimension $D_4$: Quantitative Claim Substantiation (0--3).}
$D_4$ records whether numerical performance claims are backed by source or methodology.
A score of $0$ indicates that no quantitative claim was made and is not interpreted as low or high washing.

\begin{table}[h]
\centering
\small
\caption{$D_4$: Quantitative Claim Substantiation anchors.}
\label{tab:d4_anchors}
\begin{tabularx}{\linewidth}{@{}c X@{}}
\toprule
\textbf{Score} & \textbf{Anchor description} \\
\midrule
0 & \textbf{N/A.} No specific numerical claims made. \\
\addlinespace
1 & \textbf{Unsubstantiated.} Numbers claimed with no source or methodology, including \enquote{up to X\%} framing without baseline (e.g., \enquote{4x more leads}, \enquote{saves 90\% of costs}). \\
\addlinespace
2 & \textbf{Partially Substantiated.} Numbers with some context but no detailed methodology (e.g., ROI calculators with adjustable assumptions, customer-specific outcomes without independent verification). \\
\addlinespace
3 & \textbf{Fully Substantiated.} Numbers backed by named sources, described methodology, or transparent calculations. \\
\bottomrule
\end{tabularx}
\end{table}

\paragraph{Dimension $D_5$: Marketing Proportionality (1--5).}
$D_5$ records the match between the prominence of AI in marketing and its actual role in the product.

\begin{table}[h]
\centering
\small
\caption{$D_5$: Marketing Proportionality anchors.}
\label{tab:d5_anchors}
\begin{tabularx}{\linewidth}{@{}c X@{}}
\toprule
\textbf{Score} & \textbf{Anchor description} \\
\midrule
1 & \textbf{Severely Disproportionate.} AI dominates all branding and messaging but the product has little or no detectable AI. \\
\addlinespace
2 & \textbf{Overstated.} AI heavily emphasised but plays a supporting or limited role in the product. \\
\addlinespace
3 & \textbf{Somewhat Overstated.} AI is real and prominent in the product but marketed as more advanced than the available evidence supports. \\
\addlinespace
4 & \textbf{Proportionate.} Marketing prominence roughly matches the product's actual AI capability. \\
\addlinespace
5 & \textbf{Understated or Honest.} Conservative marketing focused on outcomes, with explicit discussion of limitations where relevant. \\
\bottomrule
\end{tabularx}
\end{table}

\paragraph{Dimension $D_6$: Ethical Concern Level (0--3).}
$D_6$ records the ethical risk of the AI application itself, coded independently of marketing behaviour.
A score of $3$ was grounded in the EU AI Act's high-risk definitions and documented with the specific evidence for that startup.

\begin{table}[h]
\centering
\small
\caption{$D_6$: Ethical Concern Level anchors.}
\label{tab:d6_anchors}
\begin{tabularx}{\linewidth}{@{}c X@{}}
\toprule
\textbf{Score} & \textbf{Anchor description} \\
\midrule
0 & \textbf{No Concerns.} Benign application (e.g., e-commerce shop builder, text simplification). \\
\addlinespace
1 & \textbf{Minor.} Some potential for misuse or minor privacy implications (e.g., automated outreach, personalisation). \\
\addlinespace
2 & \textbf{Moderate.} Personal-data processing, behavioural tracking, or cross-device tracking. \\
\addlinespace
3 & \textbf{Serious.} Surveillance, emotion detection, discriminatory potential, lack of human oversight in regulated domains, or an application that meets an EU AI Act high-risk category. Annotators recorded the ethical subcategory (surveillance, privacy, manipulation, discrimination, lack of human oversight, vulnerable persons) alongside the score. \\
\bottomrule
\end{tabularx}
\end{table}

\paragraph{Dimension $D_7$: Transparency About Limitations (1--4).}
$D_7$ records whether the startup discloses what its AI cannot do, including accuracy bounds and human-in-the-loop dependencies.

\begin{table}[h]
\centering
\small
\caption{$D_7$: Transparency About Limitations anchors.}
\label{tab:d7_anchors}
\begin{tabularx}{\linewidth}{@{}c X@{}}
\toprule
\textbf{Score} & \textbf{Anchor description} \\
\midrule
1 & \textbf{No Acknowledgment.} AI presented as infallible. No caveats anywhere on the public-facing site. \\
\addlinespace
2 & \textbf{Minimal.} Brief generic disclaimers (e.g., small-print \enquote{results may vary}, \enquote{AI-generated content can be faulty}). \\
\addlinespace
3 & \textbf{Moderate.} Acknowledges boundaries or input dependence without detailed disclosure; may include self-reported accuracy without methodology. \\
\addlinespace
4 & \textbf{Comprehensive.} Explicitly states what the AI can and cannot do, with accuracy ranges or human-in-the-loop scope where applicable. \\
\bottomrule
\end{tabularx}
\end{table}

\paragraph{Composite AI-Washing Score (1--5).}
The composite is not an arithmetic mean of $D_1$--$D_7$.
It is a holistic qualitative judgment that integrates the seven dimensions, with adjustments for ethical-concern elevation and for the mitigating effect of strong evidence or disclosed limitations (see Section~\ref{sec:codebook}).

\begin{table}[h]
\centering
\small
\caption{Composite AI-Washing Score anchors.}
\label{tab:composite_anchors}
\begin{tabularx}{\linewidth}{@{}c X@{}}
\toprule
\textbf{Score} & \textbf{Anchor description} \\
\midrule
1 & \textbf{No AI-Washing.} Transparent, proportionate, evidenced, and ethically sound across the seven dimensions. \\
\addlinespace
2 & \textbf{Mild.} Minor exaggerations but fundamentally honest. Real AI with slightly overstated marketing. \\
\addlinespace
3 & \textbf{Moderate.} Notable disconnect between claims and reality across a few dimensions (e.g., an orchestrator marketed as a unified proprietary AI; a wrapper presented with sweeping autonomy claims). \\
\addlinespace
4 & \textbf{Severe.} Significant misrepresentation across multiple dimensions (e.g., vague claims combined with a wrapper backend and overstated marketing). \\
\addlinespace
5 & \textbf{Extreme.} Misrepresentation across nearly all dimensions, in some cases with ethical obfuscation. \\
\bottomrule
\end{tabularx}
\end{table}

Figure~\ref{fig:app_dim_dist} reports the score distribution on each of the seven codebook dimensions and on the composite AI-washing index for the full sample.
The composite panel (bottom right) uses a different colour to distinguish the holistic judgment from the per-dimension scores.

\section{Appendix B: Written Interview Details}
\label{appendix:Survey Details}

The full survey instrument is reproduced below.
Front matter (informed consent) and the closing section (optional contact details for respondents willing to be interviewed) are omitted.
All questions were optional.

\paragraph{Demographic information.}
\begin{enumerate}
    \item What is your age?
        \begin{itemize}
            \item Below 18 years
            \item 18--24 years
            \item 25--34 years
            \item 35--44 years
            \item 45--54 years
            \item 55--64 years
            \item 65 years and older
        \end{itemize}

    \item What is your gender?
        \begin{itemize}
            \item Man
            \item Woman
            \item Non-binary / Diverse
            \item Prefer not to say
        \end{itemize}

    \item What is your highest degree or educational qualification?
        \begin{itemize}
            \item High school
            \item Bachelor's degree
            \item Master's degree
            \item Apprenticeship
            \item Other
        \end{itemize}
        \textit{Branch (if \enquote{Other} is selected):} Please specify your highest degree or educational level (free text).
\end{enumerate}

\paragraph{Field of education and current activity.}
\begin{enumerate}
    \setcounter{enumi}{3}
    \item Which field of study did you focus on? (free text)
    \item In which field are you currently working or studying?
    If you are studying and also hold a working-student job, please provide both. (free text)
\end{enumerate}

\paragraph{Familiarity with AI technologies.}
\begin{enumerate}
    \setcounter{enumi}{5}
    \item On a scale from 1 to 5, how would you assess your familiarity with Artificial Intelligence?
    \begin{center}
    \begin{tabular}{ccccc}
    \textbf{1} & \textbf{2} & \textbf{3} & \textbf{4} & \textbf{5} \\
    Very low & & & & Very high \\
    \end{tabular}
    \end{center}

    \item Have you used or interacted with AI technologies in any context?
        \begin{itemize}
            \item Yes, in a work or professional setting
            \item Yes, in personal use
            \item Yes, in both contexts
            \item No
        \end{itemize}
        \textit{Branch (if any \enquote{Yes} option is selected):}
        Please describe your experience using AI technologies, including specific tools or applications, context of use (work, personal projects, etc.), and frequency of interaction. (free text)
\end{enumerate}

\paragraph{Perceptions of Artificial Intelligence.}
\begin{enumerate}
    \setcounter{enumi}{7}
    \item How would you describe the term \enquote{Artificial Intelligence} in your own words? (free text)

    \item When you see a product or service advertised as \enquote{AI-powered}, what is your initial reaction?
    \begin{center}
    \begin{tabular}{ccccc}
    \textbf{1} & \textbf{2} & \textbf{3} & \textbf{4} & \textbf{5} \\
    Very skeptical & & & & Very trustful \\
    \end{tabular}
    \end{center}

    \item Have you encountered German startups that advertise AI capabilities in their products or services?
        \begin{itemize}
            \item Yes
            \item No
        \end{itemize}
        \textit{Branch (if \enquote{Yes}):}
        Please describe specific examples of AI claims or promises you have seen from startups. Consider: What was promised? Which company? What product or service? Did you use it? (free text)
\end{enumerate}

\paragraph{Personal experience of AI-washing.}
\begin{enumerate}
    \setcounter{enumi}{10}
    \item Have you personally encountered what you would consider AI-washing, either as a consumer or in your work?
        \begin{itemize}
            \item Yes
            \item No
        \end{itemize}
        \textit{Branch (if \enquote{Yes}):}
        Please describe your AI-washing experience.
        Include: What was claimed? What was the reality? How did you discover the discrepancy? What were the consequences? (free text)
\end{enumerate}

\paragraph{Assessing current AI capabilities.}
For each application below, please assess the current state of AI technology.

\begin{enumerate}
    \setcounter{enumi}{11}
    \item Do you believe that AI is already capable of reliably and ethically deploying surveillance systems (e.g., video surveillance, behavioural analysis) today?
    \begin{center}
    \begin{tabular}{ccccc}
    \textbf{1} & \textbf{2} & \textbf{3} & \textbf{4} & \textbf{5} \\
    Not capable & & & & Fully capable \\
    \end{tabular}
    \end{center}

    \item Do you believe that AI is already capable of making complex, autonomous decisions (e.g., business strategy, medical diagnosis, legal judgements)?
    For example, in case of an injury, an AI system telling you what to do.
    \begin{center}
    \begin{tabular}{ccccc}
    \textbf{1} & \textbf{2} & \textbf{3} & \textbf{4} & \textbf{5} \\
    Not capable & & & & Fully capable \\
    \end{tabular}
    \end{center}

    \item Do you believe that AI is already capable of taking over the role of therapists and psychologists in addressing mental health problems?
    \begin{center}
    \begin{tabular}{ccccc}
    \textbf{1} & \textbf{2} & \textbf{3} & \textbf{4} & \textbf{5} \\
    Not capable & & & & Fully capable \\
    \end{tabular}
    \end{center}
\end{enumerate}

\paragraph{Final thoughts.}
\begin{enumerate}
    \setcounter{enumi}{14}
    \item Is there anything else you would like to share about AI-washing, AI marketing, or AI technologies in general? (free text)
\end{enumerate}

\clearpage
\section{Appendix C: Oral Interview Details}
\label{appendix:Interview Details}

Before each interview, the interviewer read the following consent script aloud and confirmed the participant's verbal agreement before starting the recording:
 
\begin{quote}
\enquote{I am author, in this study, I am running a mixed-methods study on AI-washing, which is a marketing practice where companies overstate or falsely claim the use of artificial intelligence in their products.
To increase the body of knowledge on this topic, we are conducting surveys, interviews, and a quantitative study on AI-washing.
This is the interview part, which you agreed to do with me.
The interview will last around 20--30 minutes.
I will be recording this interview, and I will discard the recordings once I finalise my notes.
In this interview, I am going to ask about your experience and opinions on AI and AI-washing.
You will be completely anonymous in our study; we will not reveal your name or affiliation.
All of the questions are optional; you can skip any question that you want to.}
\end{quote}

Semi-structured interviews followed the guide below.
All questions were optional, and the order was adapted to the flow of each conversation.

\begin{enumerate}
    \item What is your age, gender, and place of residence?
    \item What is your educational background?
    \item In which field are you currently working or studying?
    \item Have you gained any professional or private experience with AI technologies?
    If yes, which ones?
    Do you work with AI in your professional or private life?
    \item How would you describe the term \enquote{Artificial Intelligence} in your own words?
    What kinds of tasks does this imply?
    \item How would you assess your familiarity with Artificial Intelligence?
    \item What do you think the future of AI will bring in the next 5 to 10 years?
    \item What role do you think AI startups play in the development of AI technologies?
    \item Have you personally observed examples of startups promising AI applications?
    \item What is your opinion on AI startups in Germany?
    \item What thoughts do you have when you see an advertisement promoting an AI functionality?
    Do you consider those promises to be realistic?
    \item Have you ever experienced AI-washing as a consumer or in your profession?
    \item \textit{If the participant works in marketing:}
        \begin{itemize}
            \item Have you ever been involved in the advertising process for a product with AI capability?
            \item Do you feel pressure to include AI in your marketing campaigns?
            If yes, why?
        \end{itemize}
    \item Based on your experience and work, could AI solve tasks in your area?
    \item Do you believe that AI is already capable of reliably and ethically deploying surveillance systems (e.g., video surveillance) today?
    Why or why not?
    \item How do you rate the ability of AI to independently write essays, articles, or texts at a quality comparable to human authors?
    \item In your opinion, can AI systems autonomously make complex decisions?
    Please elaborate.
    \item Is there anything else you would like to cover in this interview?
\end{enumerate}


\end{document}